\documentclass{article}
\usepackage[english]{babel}
\usepackage[utf8x]{inputenc}
\usepackage{geometry}
\usepackage{indentfirst}
\usepackage{amsmath}
\usepackage{graphicx}
\usepackage{float} 
\usepackage{array}
\usepackage{multirow}
\usepackage{cite}
\usepackage{textcomp}

\title{\textbf{Effect of Electrode Array Position on Electric Field Intensity in Glioblastoma Patients Undergoing Electric Field Therapy}}

\author{
Yousun Ko\textsuperscript{1}, Sangcheol Kim\textsuperscript{1}, Tae Hyun Kim\textsuperscript{2},Dongho Shin\textsuperscript{2},
Haksoo Kim\textsuperscript{2},\\Sung Uk Lee\textsuperscript{2\textasteriskcentered},
Jonghyun Kim\textsuperscript{3}, Myonggeun Yoon\textsuperscript{1,3\textasteriskcentered} \vspace{0.5em}\\[1ex]
\textsuperscript{1}Department of Bio-medical Engineering, Korea University, Seoul, Republic of Korea\\
\textsuperscript{2}National Cancer Center, Seoul, Republic of Korea\\
\textsuperscript{3}Field Cure Ltd., Seoul, Republic of Korea
}

\date{}

\begin{document}
\maketitle

\rule{0.95\linewidth}{0.5pt}

\begin{abstract}\noindent\textbf{Background:} The intensity of the electric field applied to a brain tumor by electric field therapy is influenced by the position of the electrode array, which should be optimized based on the patient’s head shape and tumor characteristics. This study assessed the effects of varying electrode positions on electric field intensity in glioblastoma multiforme (GBM) patients.\\
\textbf{Methods:} This study enrolled 13 GBM patients. The center of the MR slice corresponding to the center of the tumor was set as the reference point for the electrodes, creating pairs of electrode arrays in the top-rear and left-right positions. Based on this reference plan, four additional treatment plans were generated by rotating three of the four electrode arrays, all except the top electrode array, by ± 15\textdegree{}\, and ± 30\textdegree{}\, from their reference positions, resulting in a total of five treatment plans per patient. Electric field frequency was set at 200 kHz, and current density at 31 mArms/cm\textsuperscript{2}. The minimum and mean electric field intensities, homogeneity index (HI), and coverage index (CovI) were calculated and compared.\\
\textbf{Results:} The optimal plans showed differences ranging from -0.39\% to 24.20\% for minimum intensity and -14.29\% to 16.67\% for mean intensity compared to reference plans. HI and CovI varied from 0.00\% to 48.65\% and 0.00\% to 95.3\%, respectively. The average improvements across all patients were 8.96\% for minimum intensity, 5.11\% for mean intensity, 15.65\% for HI, and 17.84\% for CovI.\\
\textbf{Conclusions:} Optimizing electrode angle improves electric field therapy outcomes in GBM patients by maximizing field intensity and coverage.\vspace{0.5em}\vspace{0.5em}\\
\textbf{Key Points}
\vspace{-0.5em} 

\begin{enumerate}
  \item Adjusting electrode positions enhances electric field intensity in GBM treatment.
  \item Customizing electrode positions based on individual patient anatomy enhances electric field therapy effectiveness.
\end{enumerate}

\noindent\textbf{Importance of the Study}\\
This study highlights the importance of optimizing the electrode array positions in electric field therapy for GBM patients. By using real patient data, the study demonstrates that customizing electrode array angles according to each patient’s unique cranial structure and tumor location can significantly increase the electric field intensity reaching the tumor. This approach enhances the therapeutic effectiveness by maximizing the electric field’s ability to disrupt tumor cell division, which is crucial for improving patient outcomes. The findings suggest that personalized treatment plans, tailored to the specific anatomical and tumor characteristics of each patient, are essential to achieving optimal results in electric field therapy, potentially leading to better control of tumor growth and longer survival in patients with GBM.
\end{abstract}

\rule{0.95\linewidth}{0.5pt}

\vspace{1em}

\section{Introduction}
Glioblastoma multiforme (GBM) has been classified by the World Health Organization (WHO) as a grade IV brain tumor, with a 5-year post-diagnosis survival rate below 10\%. The current standard treatment includes maximal safe resection followed by radiotherapy plus temozolomide (TMZ, 75 mg/m\textsuperscript{2}) and subsequent adjuvant TMZ (150–200 mg/m\textsuperscript{2})~\cite{ref1}. Despite these treatments, however, most GBMs recur after treatment~\cite{ref2}. Moreover, GBM is an extremely lethal cancer, with a median overall survival of 14.6 to 16.0 months and a median progression-free survival of only 4.0 months~\cite{ref3}. Electric field therapy is an emerging adjunctive treatment following surgery and chemoradiotherapy in patients newly diagnosed with GBM~\cite{ref4}. Moreover, electric field therapy has been reported to increase overall survival in patients with GBM~\cite{ref5}.

Electric field therapy consists of the application of a low-intensity (1–3 V/cm), intermediate-frequency (100–300 kHz) alternating electric field through two pairs of electrode arrays, which inhibits and disrupts the division of proliferating tumor cells~\cite{ref6, ref7, ref8}. The therapeutic effect of this treatment is dependent on the intensity of the electric field that reaches the tumor~\cite{ref9}. Insufficient electric field intensity may reduce the ability of the electric field to inhibit tumor cell division, resulting in less effective treatment. Studies have therefore sought to determine the minimum electric field intensity required to achieve successful tumor treatment, with results showing that a minimum electric field intensity of 1.06 V/cm reaching the tumor is required for therapeutic efficacy~\cite{ref10}.

Optimizing the therapeutic efficacy of electric field therapy requires treatment plans that maximize the electric field intensity applied to a tumor while minimizing the field intensity applied to surrounding healthy tissues, thereby preventing unwanted side effects~\cite{ref11}. The position of the electrode arrays must therefore be adjusted based on the location of the tumor to ensure that the maximum possible electric field is delivered to the tumor. A study using a realistic model of a human head was therefore used to test the effects of altering the angles of the electrode arrays attached to the patient’s scalp on the electric field intensity applied to the tumor~\cite{ref12}. That study found that the optimal angle of placement of electrode arrays depended on tumor location. For example, when the tumor was positioned in the left-right direction, the peak electric field intensity occurred at ± 45\textdegree{}. Additionally, electrode positioning altered the electric field intensity within a range of 1.31 to 1.95 V/cm.

Although studies have assessed methods of adjusting the intensity of electric fields applied to tumors, no study to date has evaluated actual patient data. In real patients, the locations, shapes and sized of tumors vary, as do their cranial characteristics. Consequently, even if an optimal electrode angle is determined based on the tumor’s location, application of the corresponding electrode arrangement may not be feasible due to individual cranial features of the patient. Furthermore, even if the optimal angle can be determined, the electric field may not be uniformly distributed across the tumor, or the maximum electric field may not be applied to the tumor due to its shape. The present study therefore used magnetic resonance (MR) images of actual GBM patients to measure the electric field intensity that would be attained if these patients were undergoing electric field therapy and to evaluate the efficacy of electric field therapy by measuring homogeneity index (HI) and coverage index (CovI), parameters frequently measured in patients undergoing radiation therapy ~\cite{ref13}.

\section{Materials and Methods}
\subsection{Patient data and experiment settings}  
Thirteen GBM patients were selected from Burdenko’s Glioblastoma Progression Dataset ~\cite{ref13}; their demographic and clinical characteristics are listed in Table 1. Prior to electric field therapy, each tumor and brain was divided into white matter, gray matter, CSF, bone, and external regions. The gross tumor volume (GTV) of each tumor was finalized through segmentation, with the assistance of a clinical expert. The frequency and current density for electric field therapy were set to 200 kHz and 31 mArms/cm\textsuperscript{2}, respectively (Table 1).

\renewcommand{\arraystretch}{1.2} 

\begin{table}[ht]
  \centering
  \resizebox{\textwidth}{!}{%
    \begin{tabular}{ccccccc}
      \hline
      \textbf{Number of patients} & 
      \textbf{Number of males} & 
      \textbf{Number of females} & 
      \textbf{Pathology} & 
      \textbf{Modality} & 
      \textbf{\begin{tabular}[c]{@{}c@{}}Frequency\\ (kHz)\end{tabular}} & 
      \textbf{\begin{tabular}[c]{@{}c@{}}Current density\\ (mArms/cm\textsuperscript{2})\end{tabular}} \\
      \hline
      13 & 
      5 & 
      8 & 
      \begin{tabular}[c]{@{}c@{}}high grade glioma\\ /glioblastoma\end{tabular} & 
      MR(T1) & 
      200 & 
      31.00 \\
      \hline
    \end{tabular}
  }
  \caption{Patient characteristics and treatment settings.}
  \label{tab:patient-settings}
\end{table}

\subsection{Treatment planning}  
Treatment planning for electric field therapy was performed using OncoField v1.1.0 (FieldCure, Seoul, ROK). The reference point for the electrodes was set at the center of the slice corresponding to the tumor’s centroid. From this reference point, two pairs of electrode arrays—top-rear and left-right—were generated, establishing the reference plan (0\textdegree{}\,). Subsequently, three of the electrode arrays, excluding the top one, were rotated by ± 15\textdegree{}\, and ± 30\textdegree{}\, from the reference point to create additional plans. Five treatment plans were therefore generated for each patient, the reference plan and the four rotated plans (Fig. 1).

\begin{figure}[htbp]
  \renewcommand{\figurename}{Fig.}
  \centering
  \includegraphics[width=0.58\textwidth]{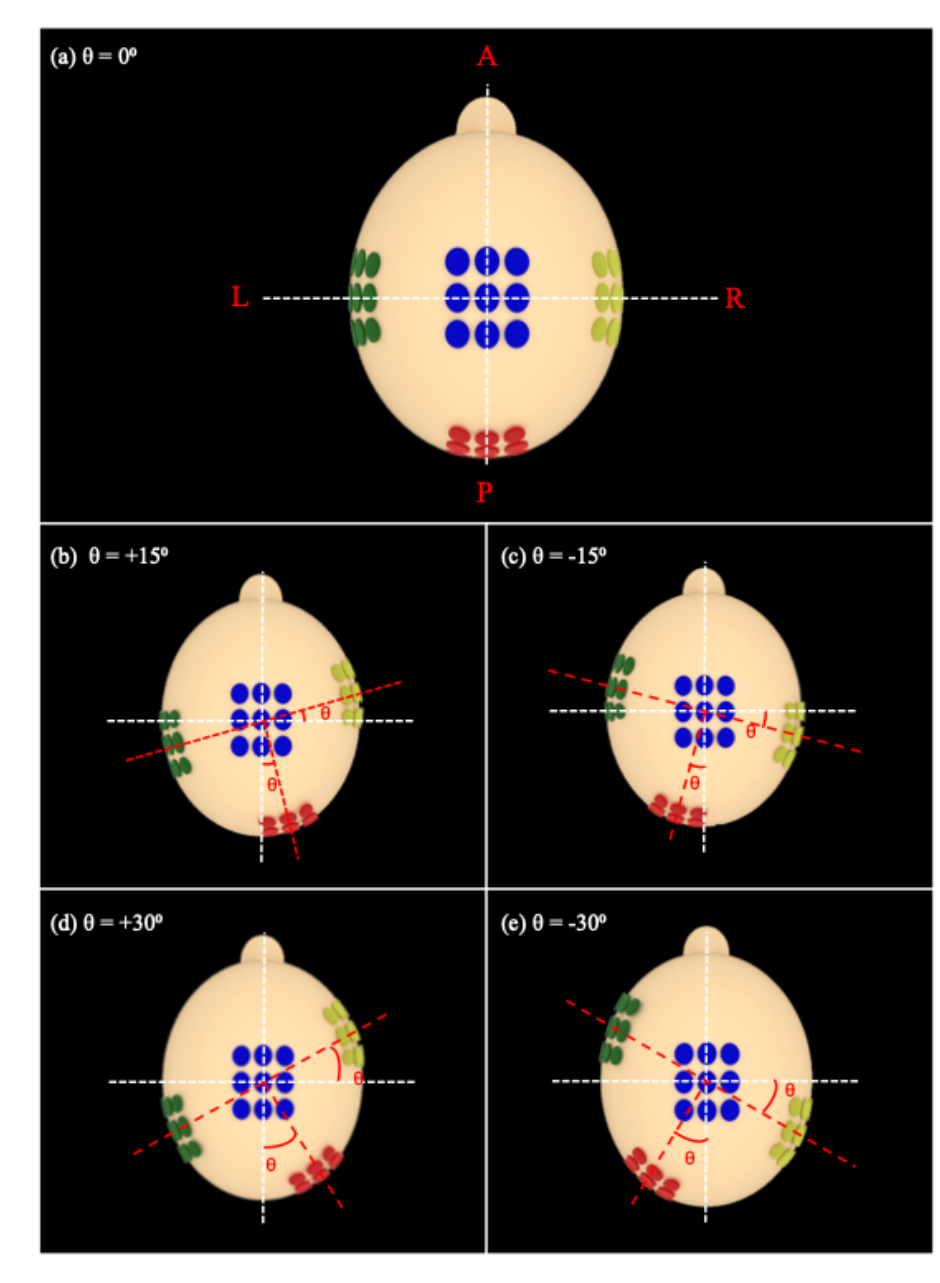}  
  \caption {Illustration of electric field treatment plans for a patient with GBM. (a) Reference plan. (b-e) The reference plan with rotation of three electrode arrays (all except the top electrode array) by (b-c) ± 15\textdegree{}\, rotation and (d-e) ± 30\textdegree{}\, rotation}
\end{figure}

\subsection{Evaluations}
The minimum, maximum, and mean electric field intensities were measured for each treatment plan. Based on these measured values, the homogeneity index (HI) was calculated using the equation ~\cite{ref13}: 

\begin{equation}
\text{HI} = \frac{D_{\text{max}} - D_{\text{min}}}{D_p}
\label{eq:HI}
\end{equation}

\noindent where $D_p$ is the prescribed electric field intensity, set to 1.06 V/cm ~\cite{ref10}.
The coverage index (CovI) was calculated using the equation:

\begin{equation}
\text{CovI} = \frac{V_{100\text{GTV}}}{V_{\text{GTV}}}
\label{eq:CovI}
\end{equation}

\noindent where $D_{100GTV}$ represents the volume within the GTV receiving at least 100\% of $D_{p}$, and $V_{GTV}$ is the total volume of the GTV. The plan with the highest minimum and mean electric field intensities for each patient was selected as the best plan for that patient. The HI was normalized to the reference plan value, with the plan having the lowest HI selected as the best plan. Similarly, the CovI was normalized to the reference plan, and the plan with the highest CovI was chosen as the best plan. The uniformity of the electric field intensity within the GTV for the five generated plans was compared using a dose-volume histogram (DVH), which is commonly used in radiation therapy to compare competing treatment plans for specific patients~\cite{ref14}.

\section{Results}
The centroid of the tumor was set as the reference point for electrode placement. Table 2 compares the reference plan with the four rotated plans across various indices. The best of the four rotated plans was selected based on previously described criteria. The values in the table represent the percentage difference between the reference plan and the best plan for each index. A negative value indicates that the reference plan performed better, while a value of zero denotes no difference. As determined by minimum electric field intensity, the rotated plans outperformed the reference plans in all patients except for patient 4. Similarly, as determined by mean electric field intensity index, the rotated plans yielded higher electric field intensities than the reference plans in all patients, except for patient 8. Evaluation of two other indices, HI and CovI, showed that the rotated plans were equal to or greater than the reference plan in all 13 patients.

\renewcommand{\arraystretch}{1.5}  

\begin{table}[ht]
  \centering
  \begin{minipage}{\textwidth}
    \centering
    \resizebox{\textwidth}{!}{%
      \begin{tabular}{c c c c c c c c c c c c c c c}
      \hline
      \begin{tabular}[c]{@{}c@{}}Patient number\end{tabular} & 
      1 & 2 & 3 & 4 & 5 & 6 & 7 & 8 & 9 & 10 & 11 & 12 & 13 & 
      \multirow{2}{*}{\begin{tabular}[c]{@{}c@{}}Average\\ improvement (\%)\end{tabular}} \\
      \cline{1-14}
      \multicolumn{14}{c}{{Reference vs. best plan for each indices}} \\
      \hline
      Min field (\%)  & 5.82  & 8.43  & 9.66  & -0.39 & 4.07  & 1.99  & 9.63  & 8.85   & 1.67  & 18.99 & 16.82 & 6.67  & 24.20 & 8.96 \\
      Mean field (\%) & 0.00  & 16.67 & 9.09  & 0.00  & 0.00  & 7.14  & 8.33  & -14.29 & 4.55  & 6.67  & 10.00 & 7.14  & 11.11 & 5.11 \\
      HI (\%)         & 23.57 & 45.20 & 7.06  & 3.29  & 0.00  & 4.98  & 0.71  & 11.99  & 48.65 & 8.36  & 5.94  & 5.93  & 37.81 & 15.65 \\
      CovI (\%)       & 26.80 & 17.85 & 13.83 & 0.00  & 0.17  & 0.33  & 0.03  & 45.32  & 0.17  & 7.76  & 95.26 & 20.32 & 4.08  & 17.84 \\
      \hline
    \end{tabular}
  }\\[0.4ex]  % ← Line break and add space here!
    \noindent\hfill\textit{\footnotesize Abbreviations: HI = homogeneity index; CovI = coverage index.}

    \caption{
      Percentage differences between reference plans and best plans for each index in all 13 patients,\\
      and average improvements for each index.
    }
    \label{tab:index-comparison}
  \end{minipage}
\end{table}

Two of the 13 patients were selected as representative patients. Fig. 2 shows the electric field intensity distribution for the first representative patient across different plans, showing how the electric field intensity within the tumor depended on the positions of the electrode arrays. Positive rotations of the electrode arrays resulted in a higher electric field intensity within the tumor than negative rotations. Specifically, the +30\textdegree{}\, rotation produced the best plan, whereas the -30\textdegree{}\, rotation resulted in the worst plan.

\begin{figure}[htbp]
  \renewcommand{\figurename}{Fig.}
  \centering
  \includegraphics[width=0.7\textwidth]{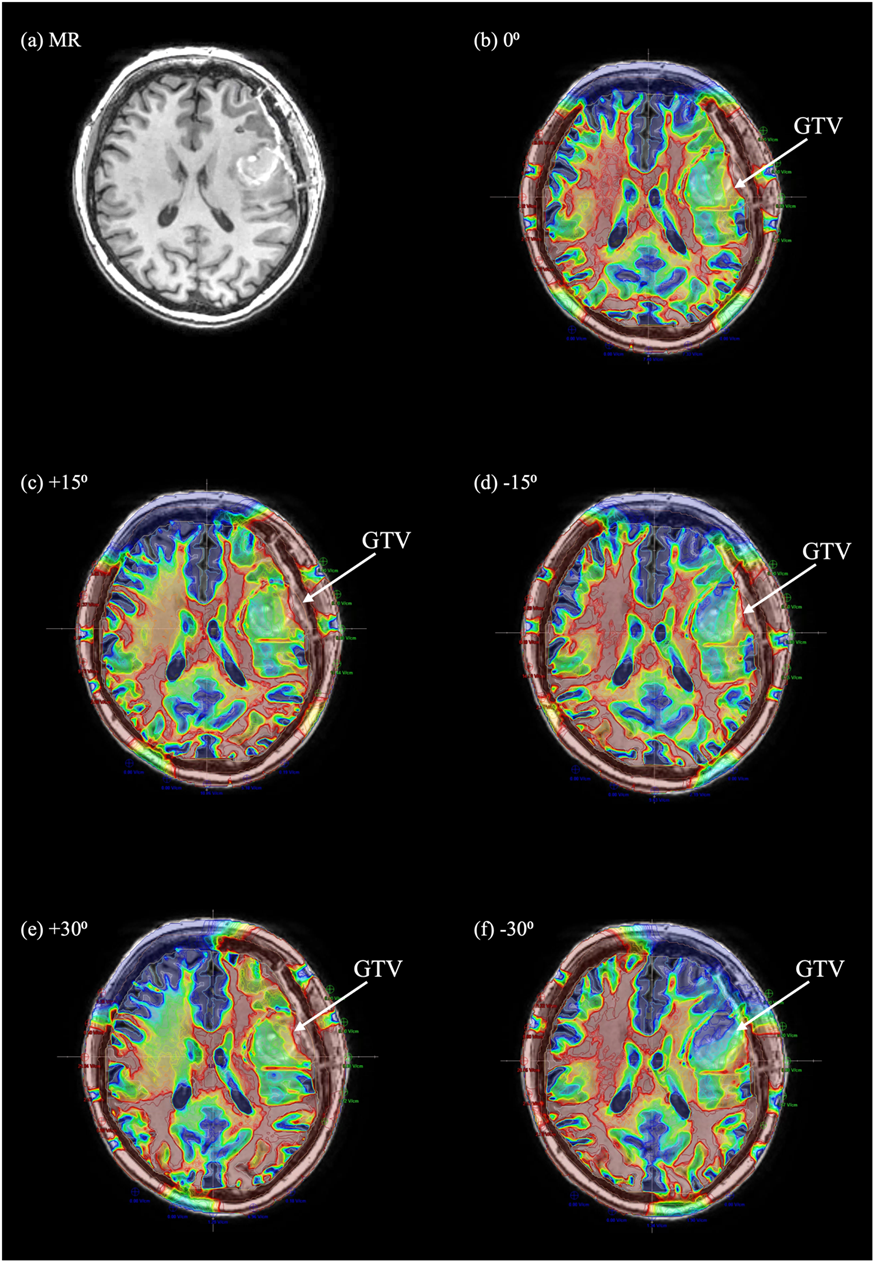}  % Include file extension
  \caption {Electric field distributions for each angle in the first representative patient}
\end{figure}

\newpage Fig. 3 presents the electric field intensity distribution for the second representative patient. In contrast to the first patient, the second patient exhibited a tendency for higher electric field intensities within the tumor when the electrode arrays were rotated in the negative than the positive direction, with the -30\textdegree{}\, rotation was identified as the best plan and the +30\textdegree{}\,  rotation as the worst plan. The tumor positions in the first and second patients were on the left and right sides, respectively, resulting in opposite trends.

\begin{figure}[htbp]
  \renewcommand{\figurename}{Fig.}
  \centering
  \includegraphics[width=0.7\textwidth]{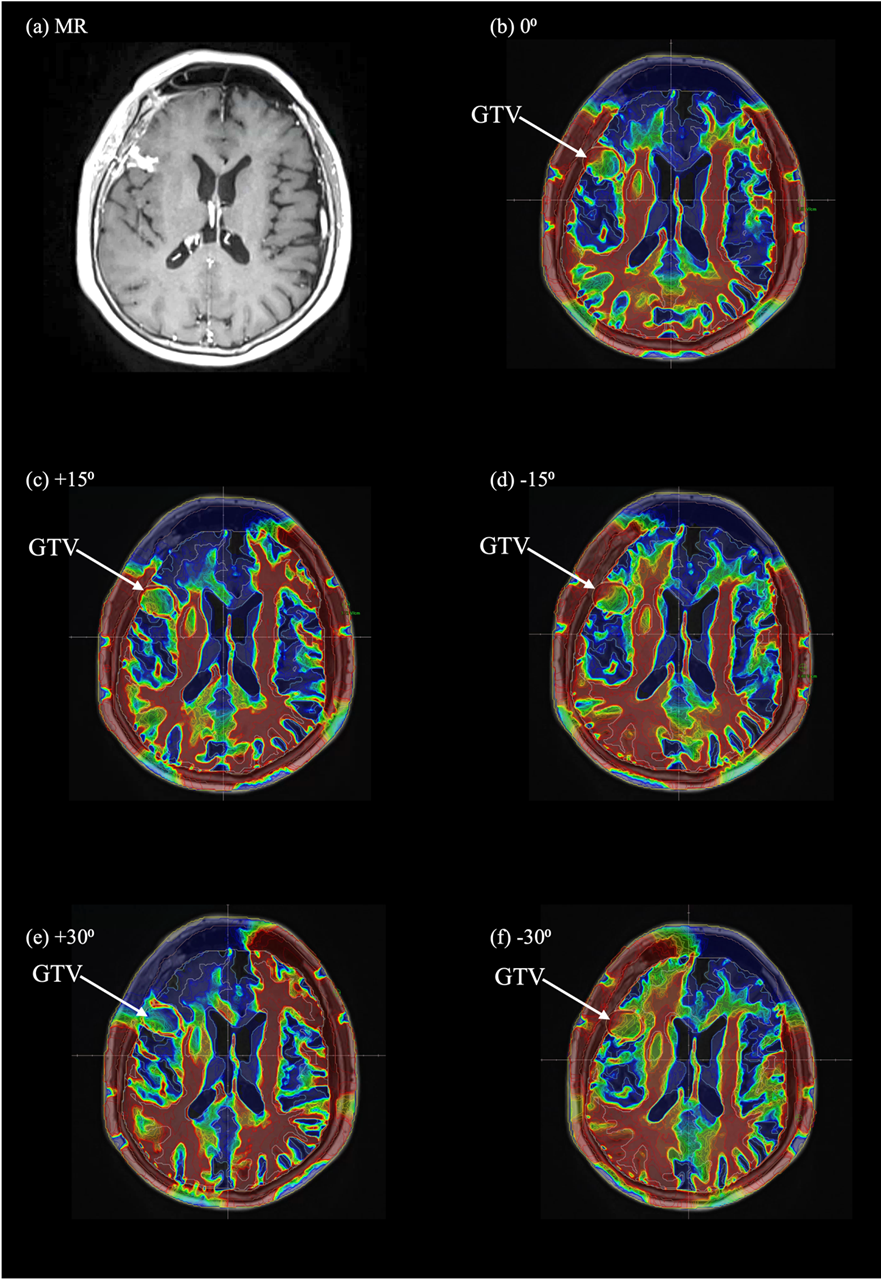}  
  \caption {Electric field distributions for each angle in the second representative patient}
\end{figure}

\newpage Fig. 4 shows the DVH for the GTV of the two representative patients. The DVH for the first representative patient confirmed that positive rotation resulted in a greater electric field intensity reaching the tumor (Fig. 4(a)), whereas the DVH for the second representative patient demonstrated that negative rotation led to a higher electric field intensity within the tumor (Fig. 4(b)).

\begin{figure}[htbp]
  \renewcommand{\figurename}{Fig.}
  \centering
  \includegraphics[width=0.7\textwidth]{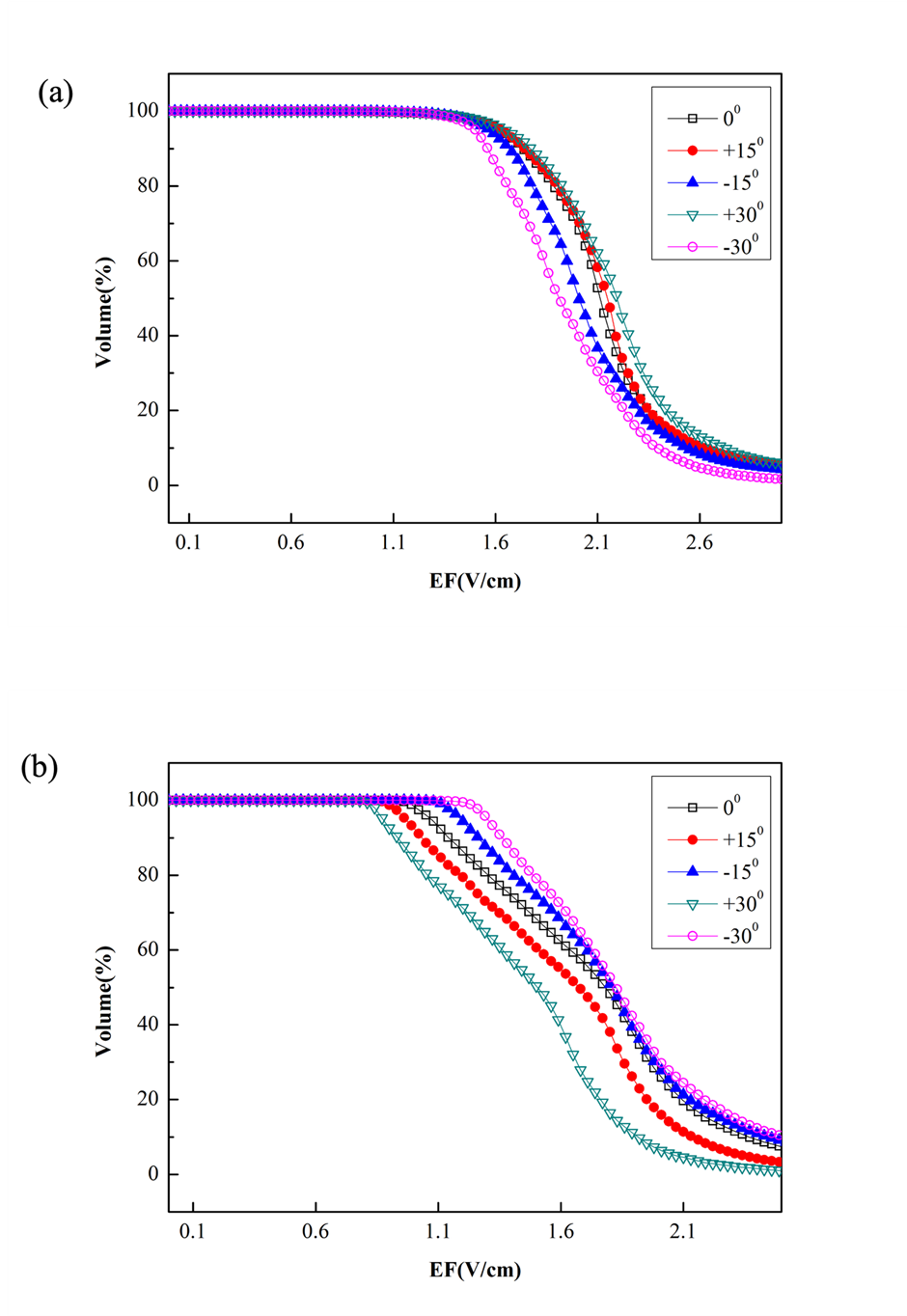} 
  \caption {(a) The DVH of the first representative patient, (b) the DVH of the second one}
\end{figure}

The last column in Table 2 presents the average percentage differences between the reference plan and the best plan for each of the four indices, as calculated from Table 1. Compared with the reference plans, the minimum field intensity for the best plan improved by 8.96\%, and the mean field intensity for the best plan improved by 5.11\%. Similarly, HI and CovI showed improvements of 15.65\% and 17.84\%, respectively.

\section{Discussion}
The present study tested the effects of electrode array angles during electric field therapy planning on changes in electric field intensity in patients with GBM. Electrode array positioning was found to significantly affect the intensity of the electric field applied to the tumor through electric field therapy. The optimal electrode positions for maximizing therapeutic efficacy depended on the cranial structure of each patient and the location and size of each tumor. Therefore, this study evaluated the effects of altering electrode array positions on changes in electric field intensity applied to the tumor in 13 actual GBM patients. Each reference plan was determined by setting the center of the MR slice corresponding to the centroid of each tumor as the reference point for the electrodes, followed by generation of top-rear and left-right electrode array pairs. Subsequently, three of these electrode arrays, excluding the top one, were rotated by ± 15\textdegree{}\, and ± 30\textdegree{}\, from the reference point, resulting in a total of five treatment plans per patient. The frequency of electric field treatment was set at 200 kHz and the current density at 31 mArms/cm\textsuperscript{2}. The minimum and mean electric field intensities, as well as HI and CovI, were calculated for each treatment plan and compared with the reference plan for that patient. Compared with the reference plans, the best of the rotated plans for each of the 13 patients showed improvements in minimum and mean electric field intensities of 8.96\% and 5.11\%, respectively, and improvements in HI and CovI of 15.65\% and 17.84\%, respectively. Thus, optimizing electrode array angles based on each patient’s cranial structure and tumor location and size can maximize the electric field intensity reaching the tumor, enhancing the therapeutic effect of treatment.

A previous study analyzed the effects of electrode array angles on electric field intensity during application of electric field therapy to artificially generated GBMs on MR images of healthy patients~\cite{ref11}. In that study, tumor size and shape remained constant, while the position of the tumor was shifted along the x-axis (left-right direction) and y-axis (anterior-posterior direction). The electrode array consisted of two pairs of orthogonal electrodes, which were rotated at 15\textdegree{}\, intervals to create different plans. When the tumor was moved along the x-axis, the highest electric field intensity (1.65–1.95 V/cm) was observed at ± 45\textdegree{}. When the tumor was shifted along the y-axis, higher electric field intensities were recorded when $y > 0$ mm and the angle was 60\textdegree{}\, or greater, with a general difference of 10–17\% between the lowest and highest values. 

Similarly, the present study used two pairs of electrode arrays, although left-right pairs were used in place of orthogonal pairs. In addition, this study rotated three electrode arrays, all except the top one, at 15\textdegree{}\, intervals. In contrast to the artificial model, however, the variation in cranial shapes among actual patients limited electrode placement, as electrodes may have been obstructed by ears or eyes. To ensure consistent comparison across all 13 patients, planning was limited to rotations of ± 15\textdegree{}\, and ± 30\textdegree{}\,, which could be applied to all patients. In contrast to the previous study, the present study found no consistent trends between tumor locations and electrode rotation angles. The tumors in real patients are not perfect spheres, and their sizes and shapes vary significantly. Therefore, even when tumors are located in the same position, the optimal electrode array angles may differ from patient to patient due to variations in tumor shapes and sizes. Furthermore, the best plan angles differed for each of the four indices assessed, making it crucial during treatment planning to set indices according to specific prioritized criteria. Additionally, in some patients, the reference plans performed best, as determined by minimum or mean electric field intensity. Optimization of treatment therefore requires comprehensive planning that considers a patient’s cranial structure, as well as tumor location, size, and shape.

The present study had several limitations. Electric field therapy was planned for only 13 patients, making it difficult to identify general trends. Additionally, because the same current density was applied to all electrodes, there were no considerations of weighting based on the distance from the tumor. Effective treatment of GBM while minimizing side effects requires maximization of the electric field intensity at the tumor site and minimization of intensity to surrounding healthy tissues. Therefore, optimizing not only the positions of the electrode arrays but the electric potential applied to each is necessary to achieve a more uniform and higher electric field within the tumor ~\cite{ref15}. While the electrode reference point in this study was defined as the tumor centroid on the central MR slice, future studies should explore more personalized approaches that account for patient-specific cranial anatomy and tumor morphology. Additional studies are needed to determine whether altering electrode array angles and varying the electric potential at each electrode can achieve a more conformal electric field distribution within the tumor, while reducing the electric field in non-tumor regions.

In summary, this study entailed creating a reference plan for each patient and subsequently generating additional plans with ± 15\textdegree{}\, and ± 30\textdegree{}\, rotations. The electric field intensity within the tumor was compared across these plans. The findings revealed that even for tumors at the same position, variations in tumor shape necessitate changes in the optimal electrode angle. Additionally, differences in patients’ cranial structures can limit achievable electrode array angles. Comprehensive consideration of these factors and adjustments of electrode array angles accordingly may enhance electric field intensity within the tumor, potentially improving the effectiveness of brain tumor treatment.

\section*{Acknowledgements}
\noindent This work was supported by National Research Foundation of Korea (NRF) grants funded by the Korean government (MSIT) (Grant Nos. NRF-2021R1A2C2008695). This work was also supported by a Korea Medical Device Development Fund grant funded by the Korean Government (the Ministry of Science and ICT, the Ministry of Trade, Industry and Energy, the Ministry of Health \& Welfare, and the Ministry of Food and Drug Safety) (Project No.1711196423, RS-2023-00254868).the Ministry of Trade, Industry and Energy(MOTIE) and Korea Institute for Advancement of Technology(KIAT) through the International Cooperative R\&D program (Project No. P0019304)

\section*{Author Contributions}
\noindent Study conception and design: Sung Uk Lee, Myonggeun Yoon. Development of methodology: Tae Hyun Kim, Dong Ho Shin, Haksoo Kim. Data acquisition, analysis, interpretation, visualization: Yousun Ko, Sangcheol Kim. All authors had full access to the data included in this manuscript and approved submission.

\section*{Conflict of Interest}
\noindent The authors declare no competing interests

\newpage
\bibliographystyle{unsrt} 
\bibliography{references}  

\end{document}